\documentclass[reprint,twocolumn,showpacs]{revtex4}
\usepackage{amsmath}
\usepackage{amssymb}
\usepackage{color}
\usepackage[pdftex]{graphicx}
\usepackage{hyperref}
\usepackage[caption=false]{subfig}
\usepackage{hyperref}
\begin{document}
\title{Large effective three-body interaction in a double-well optical lattice}
\author{Saurabh Paul}
\affiliation{Joint Center for Quantum Information and Computer Science, Joint Quantum Institute and University of Maryland, Maryland 20742, USA}
\author{Eite Tiesinga}
\affiliation{Joint Quantum Institute and Joint Center for Quantum Information and Computer Science, National Institute of Standards and Technology and University of Maryland, Gaithersburg, Maryland 20899, USA}
\date{\today}
\begin{abstract}
  We study ultracold atoms in an  optical lattice with two local minima per unit cell and show that the low energy states of a multi-band Bose-Hubbard (BH) Hamiltonian with only pair-wise interactions is equivalent to an effective single-band Hamiltonian with strong three-body interactions. We focus on a double-well optical lattice with a symmetric double well along the $x$ axis and single well structure along the perpendicular directions. Tunneling and two-body interaction energies are obtained from an exact band-structure calculation and numerically-constructed Wannier functions in order to construct a BH Hamiltonian spanning the lowest two bands. Our effective Hamiltonian is constructed from the ground state of the $N$-atom Hamiltonian for each unit cell obtained within the subspace spanned by the Wannier functions of two lowest bands. The model includes hopping between ground states of neighboring unit cells. We show that such an effective Hamiltonian has strong three-body interactions that can be easily tuned by changing the lattice parameters. Finally, relying on numerical mean-field simulations,
we show that the effective Hamiltonian is an excellent approximation of the two-band BH Hamiltonian over a wide range of lattice parameters, both in the superfluid and Mott insulator regions.
\end{abstract}
\maketitle

\begin{section}{Introduction}
  Experiments with ultracold atoms in optical lattices have been used to realize the Bose-Hubbard (BH) model \cite{jaksch_cold_1998}. Its interaction driven quantum phase transition was first observed in \cite{greiner_quantum_2002}. Ultracold atoms in optical lattices are highly tunable systems and are now increasingly used to simulate other quantum many-body Hamiltonians \cite{bloch_many-body_2008, jaksch_cold_2005}. In recent years, Hamiltonians with complex band structure have been experimentally realized, such as double-well lattices \cite{sebby-strabley_lattice_2006, lee_sublattice_2007, trotzky_time-resolved_2008, atala_direct_2013}, honeycomb, triangular, one-dimensional stripe and Kagome lattices \cite{tarruell_creating_2012,jo_ultracold_2012}, and artificial graphene \cite{uehlinger_artificial_2013}. These experiments have extended the scope of the BH model to include excited bands and richer on-site interactions. Such Hamiltonians are being used to study collective phenomena ranging from modified quantum phases \cite{wu_quantum_2006,danshita_quantum_2007,zhou_condensates_2011}, topological matter \cite{tarruell_creating_2012,atala_direct_2013,olschlager_topologically_2012,sun_topological_2011}, and to the formation of long-lived unconventional Bose-Einstein condensates in excited bands, which were theoretically studied in  Refs.~\cite{isacsson_multiflavor_2005,stojanovic_incommensurate_2008,wu_unconventional_2009,paul_formation_2013} and observed by Refs~\cite{wirth_evidence_2010,olschlager_unconventional_2011}. 

Recent studies have shown that couplings between adjacent unit cells due to atom-atom interactions are also important and can lead to density-induced tunneling. For cubic lattices, such additional terms in the BH Hamiltonian have been shown to measurably modify the location of the superfluid to Mott phase boundaries \cite{luhmann_multi-orbital_2012,jurgensen_density-induced_2012}. 

Many-body Hamiltonians with three-body interaction are also being explored. Three-body interactions can lead to significantly modified quantum phases \cite{daley_atomic_2009,sowinski_exact_2012,safavi-naini_first-order_2012}, and can yield Pfaffian-like ground states \cite{greiter_paired_1991,wojs_global_2010,paredes_pfaffian-like_2007,roncaglia_pfaffian_2010} that have promising uses in quantum computation \cite{Kitaev20032}. Following the proposals by Refs.~\cite{johnson_effective_2009,tiesinga_collapse_2011,bissbort_effective_2012,mahmud_dynamics_2013}, the first experiments confirmed the presence of weak effective three-body interactions in optical lattices \cite{campbell_imaging_2006,will_time-resolved_2010,mark_precision_2011,mark_preparation_2012}. These effective three-body interactions, due to virtual transition of atoms to higher bands, are smaller than the usual two-body interaction.  Recent proposals have investigated ways to reduce the strength of this two-body term, thereby enhancing the role of three-body interactions. In fact, Ref.~\cite{daley_effective_2014} by driving the lattice at rf frequencies and Ref.~\cite{petrov_three-body_2014} by adding resonant radiation to couple internal states of an atom have proposed ways to turn off the two-body interaction altogether. It was shown in \cite{hafezi_engineering_2014} that arrays of superconducting Josephson junctions can also mimic strong three-body interaction.

  In this paper we show that very strong effective three-body interactions can be created using optical lattice potentials with two local minima per unit cell. We study trapped atoms in a double-well optical lattice with three or less atoms per site, and show that over a wide range of lattice parameters, the on-site interactions can be described by an effective Hamiltonian of the form
  \begin{align}\label{eq:modelH}
    H_{\rm eff}&=\sum_{m=1}^3\frac{1}{m!}\Gamma_m b^{\dagger m}b^m,
  \end{align}
where $b^{\dagger}(b)$ are creation(annihilation) operators, and $\Gamma_m$ represents the effective $m$-body interaction energy. We show that tuning the lattice parameters can create situations when $\Gamma_3\gg\Gamma_2$, i.e., a system with a strong effective three-body interaction energy. 

The remainder of the paper is setup as follows. In Sec.~\ref{sec:Hamiltonian} we introduce the double-well optical lattice potential and numerically obtain the band structure and Wannier functions. We then introduce the multi-band BH Hamiltonian and calculate hopping and interactions parameters of this Hamiltonian. In Sec.~\ref{sec:numericalMF} we perform a mean-field calculation to determine its phase diagram and discuss the Superfluid-Mott insulator transition. The effective Hamiltonian model is introduced in Sec.~\ref{sec:Heff} and constructed in a three-step process. In Sec.~\ref{sec:MPE} we discuss the initial step of obtaining a many-particle basis  and energy levels for each unit cell. The effective Hamiltonian in a unit cell of the form \eqref{eq:modelH} is determined in Sec.~\ref{sec:effHint}. We also show how $\Gamma_3/\Gamma_2$ can be optimized by tuning the lattice parameters. In Sec.~\ref{sec:effHhop} we study the coupling between the many-particle states in adjacent unit cells and introduce the effective tunneling Hamiltonian. We discuss the validity of the effective Hamiltonian by comparing the numerical mean-field results obtained for the effective Hamiltonian picture and the full BH Hamiltonian in Sec.~\ref{sec:compareH}. Finally, we summarize our results and present conclusions in Sec.~\ref{sec:conclusion}.
\end{section}

\begin{section}{hamiltonian for the double-well optical lattice}\label{sec:Hamiltonian}

  \begin{figure}
    \centering
    \subfloat[Part 1][]{\includegraphics[width=1.85in]{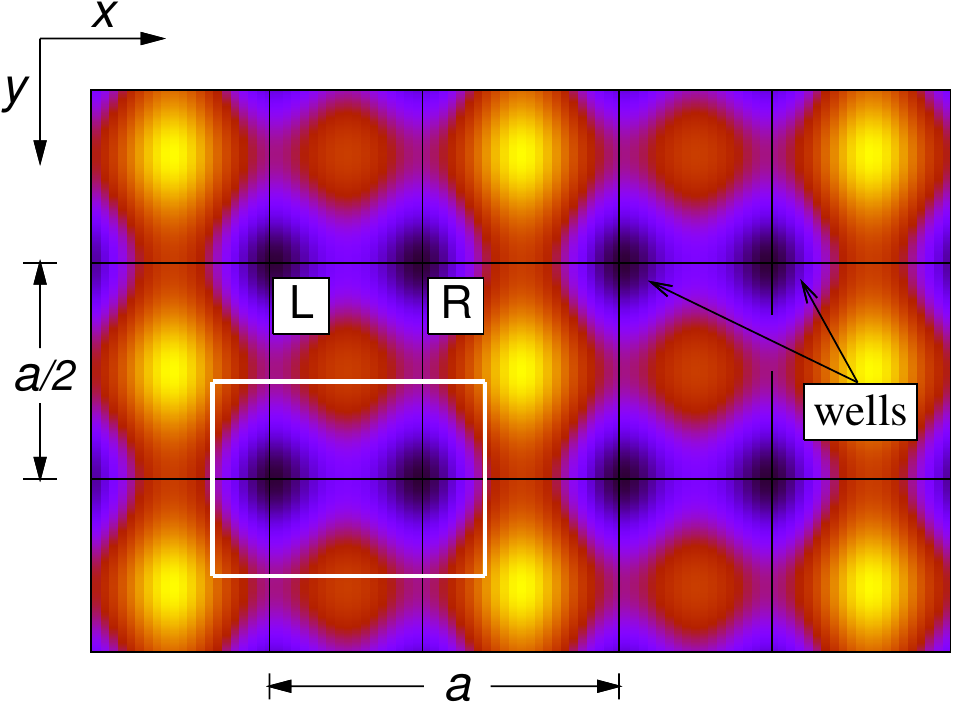}\label{fig:contour}}
    \hspace{2mm}
    \subfloat[Part 2][]{\includegraphics[width=1.35in]{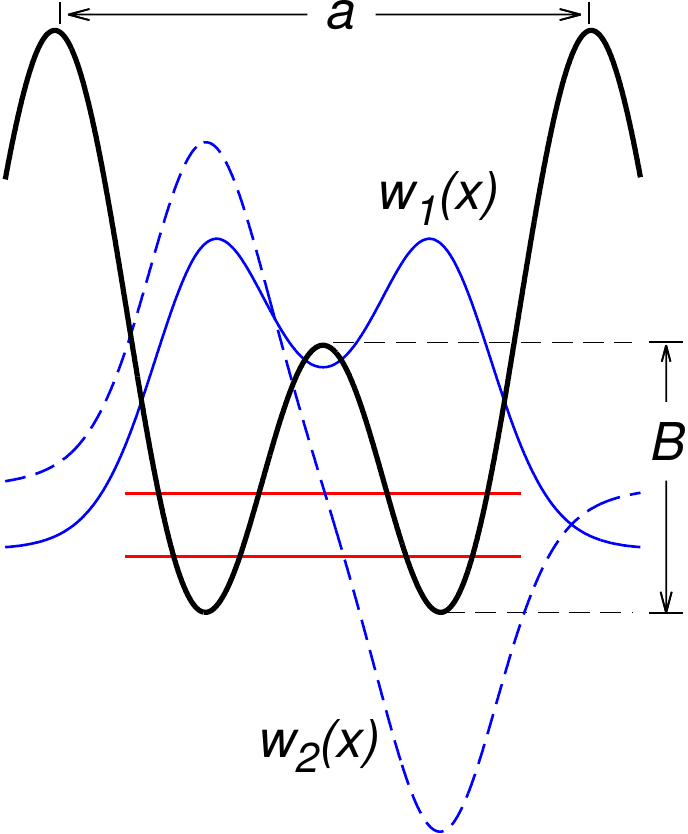}\label{fig:double-well}}
    \caption{(color online) (a) Contour plot of the optical lattice potential in the $xy$ plane, where the potential minima are in dark blue. The white box encloses a unit cell of length $a$ and $a/2$ along $x$ and $y$, respectively. Each unit cell has a double well along the $x$ axis, labeled $L$ and $R$, and a single well along the $y$ and $z$ axes. (b) Schematic of a symmetric double-well potential (thick solid black curve) in a unit cell along the $x$ axis.  The barrier height between the left and right wells is $B$. The Wannier functions for the ground and first-excited band, $w_1(x)$ and $w_2(x)$, are shown as blue curves. They are computed for a lattice with $V_0/E_R=17.0$, $V_1/V_0=1.3$, $V_2/E_R=40.0$ and $bk_L=\pi/4$. The two horizontal red curves represent their energies with splitting $\delta$.}
  \end{figure}

\begin{figure}
  \subfloat{\includegraphics[width=3.11in]{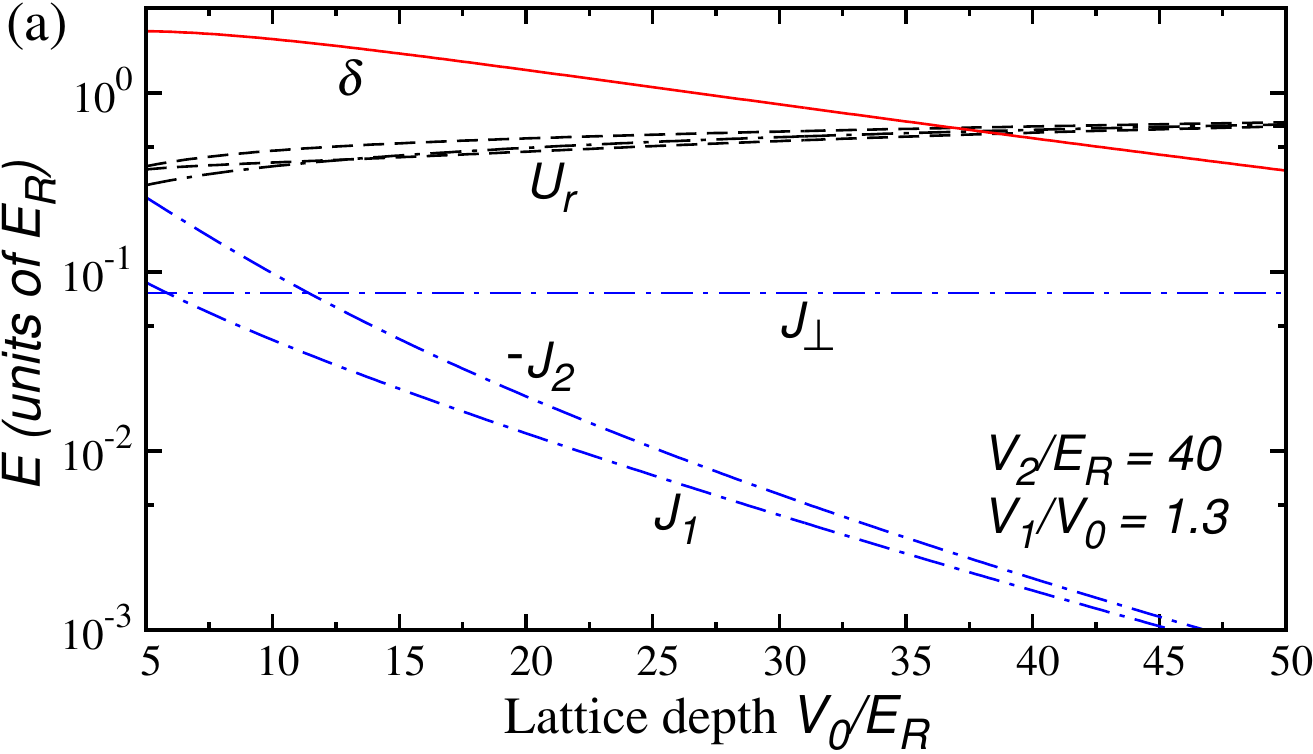}\label{fig:BHparameters}}
  \hspace{9mm}
  \subfloat{\includegraphics[width=3.2in]{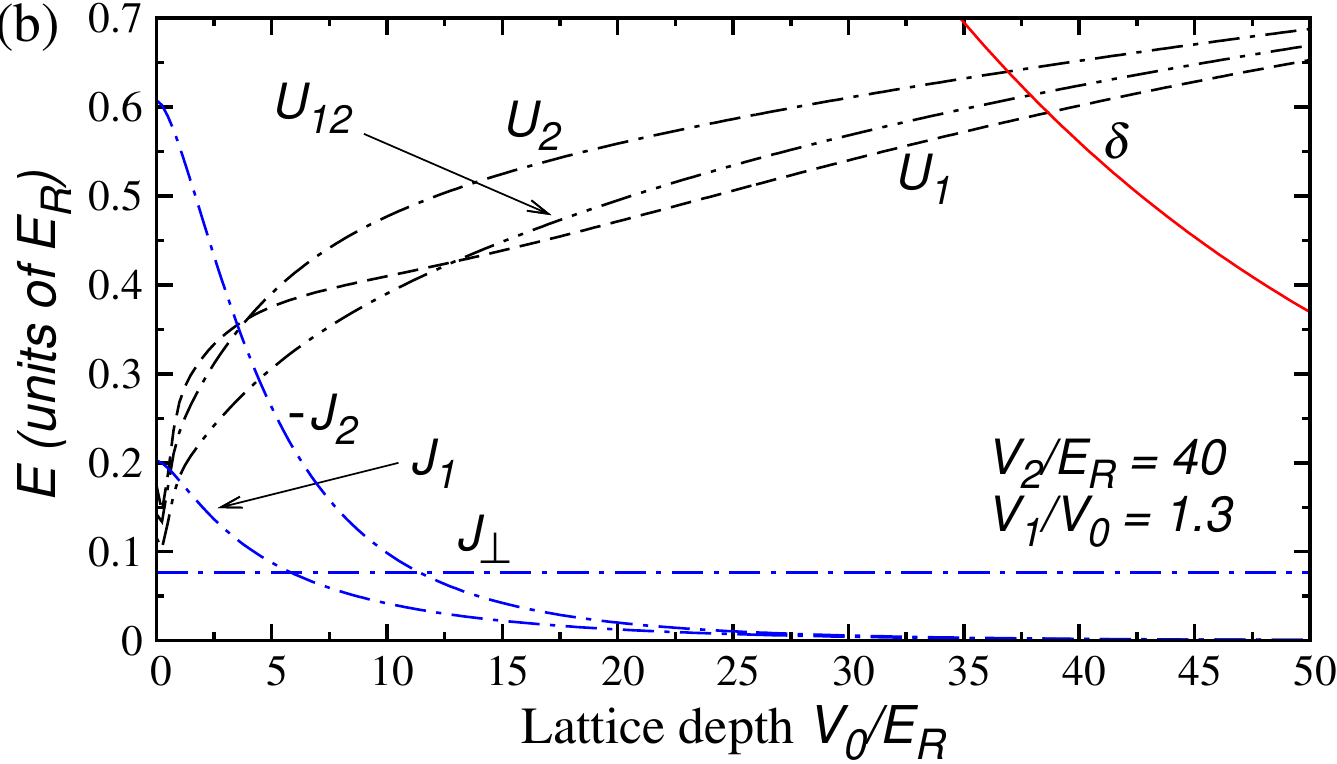}\label{fig:bandwannier}}
  \caption{(color online) Numerical parameters for the BH Hamiltonian for ${}^{87}{\rm Rb}$ on a log-linear scale (a) and linear-linear scale (b) plotted as functions of lattice depth $V_0$ in units of $E_R$, at $V_1/V_0=1.3$, $V_2/E_R=40.0$ and $bk_L=\pi/4$. The tunneling energies $J_1,-J_2$ and $J_{\perp}$ are shown as dot-dashed blue curves. The band-gap $\delta$ is plotted as a solid red curve. The interaction energies $U_r=U_1,U_2$ and $U_{12}$ are shown as dashed black curves. They are calculated for a scattering length of $a_s=100a_0$, where $a_0=0.0529\rm\,nm$ is the Bohr radius.}
\end{figure}

  We are interested in atoms in optical lattice potentials that have two nearly degenerate local minima in an unit cell. In particular, we focus on a lattice with a double-well structure along the $x$ axis and single-well structure along the perpendicular $y$ and $z$ axes. Such a lattice can be constructed by using a laser with wave vector $k_L$ and its first higher harmonic. We focus on
  \begin{align}
    \label{eq:potential}
    V(\vec x)&=-V_0\cos^2(k_Lx)-V_1\cos^2\left[2k_L(x+b)\right]\nonumber\\
        &\quad\,-V_2\left[\cos^2(2k_Ly)+\cos^2(2k_Lz)\right],
  \end{align}
where $V_{0,1,2}$ are lattice depths and $b$ is a parameter that determines whether the lattice has a symmetric or asymmetric double-well. The lattice has a periodicity of $a=\pi/k_L$ along $x$ axis, and $a/2$ along the perpendicular directions. Throughout, we express energies in units of the recoil energy along the $x$ axis, $E_R=\hbar^2k_L^2/2m_a$, where $m_a$ is the atomic mass. We note that the recoil energy along the perpendicular directions is $E_{R,\perp}=4E_R$. 
Figure~\ref{fig:contour} shows a contour plot of the optical-lattice potential in the $xy$ plane, while Fig.~\ref{fig:double-well} shows a symmetric double-well for $k_Lb=\pi/4$ along the $x$ axis.
  
  In this paper, we solely focus on the symmetric double-well lattice with $k_Lb=\pi/4$ and allow the lattice depths to vary. The barrier height between the left and right wells is $B=(V_1-V_0/4)^2/V_1$. The barrier disappears when $V_1/V_0<1/4$. In the limit $V_1/V_0\to \infty$, the potential approaches a single-well potential with period $a/2$ along all directions. Some remarks about the case $k_Lb\ne\pi/4$ are made at the end of Sec.~\ref{sec:effHint}.

  The band structure and corresponding Bloch functions are obtained numerically. They are independently computed for the $x, y$ and $z$ directions and lead to a band dispersion $\epsilon_{\vec \alpha}(\vec k)=\epsilon_{x,\alpha_x}(k_x)+\epsilon_{y,\alpha_y}(k_y)+\epsilon_{z,\alpha_z}(k_z)$, where $\vec \alpha=(\alpha_x,\alpha_y,\alpha_z)$ are the band indices and $\vec k$ is the quasi-momentum in the first Brillouin zone. The lowest two bands, $\alpha_x=1,2$, along the $x$ axis are quasi-degenerate for deep lattices. Along the perpendicular directions, the ground band $\alpha_y=\alpha_z=1$ is far removed from the first excited band. The Wannier functions, labeled by unit cell ${\bf i}=(i_x,i_y,i_z)$ and $\vec \alpha$, are a product of one dimensional Wannier functions, one for each axis, i.e., $w_{{\bf i},\vec \alpha}(\vec x)=w_{{\bf i},\alpha_x}(x)w_{{\bf i},\alpha_y}(y)w_{{\bf i},\alpha_z}(z)$. These functions can be obtained with the Marzari-Vanderbilt scheme \cite{marzari_maximally_2012} of constructing maximally localized Wannier functions. This was used for a double-well optical lattice in Ref.~\cite{modugno_maximally_2012}. Here, we follow Refs.~\cite{kivelson_wannier_1982,uehlinger_artificial_2013} and Wannier functions are found as eigen states of the position operators $\hat x$, $\hat y$ and $\hat z$ within the subspace of all the Bloch functions with band index $\vec \alpha$. To ensure real-valued Wannier functions, a real-valued discrete variable representation with periodic boundary conditions \cite{colbert_novel_1992} over $\vec M=(M_x,M_y,M_z)$ unit cells is used. Figure~\ref{fig:double-well} shows the Wannier functions $w_{{\bf i},\alpha_x}(x)$ for the lowest two bands along the $x$ axis for a symmetric lattice. We see that the Wannier functions are exponentially suppressed between unit cells, but that they are extended over the barrier between the left and right well. This is typical for the lattice parameters that are the focus of this paper.

Since the lowest two bands along $x$ axis are quasi-degenerate for deep lattices, we construct a multi-band Bose-Hubbard (BH) Hamiltonian $H$ spanning the lowest two bands along the $x$ axis, and only the ground band along the perpendicular directions. For simplicity, we drop the band indices $\alpha_y=\alpha_z=1$, and use $\alpha=1,2$ to refer to the band index along the $x$ axis. We find $H=\sum_{\bf i}\left\{H_{\bf i}^{\rm hop}+H_{\bf i}^{\rm cell}\right\}$, where the nearest-neighbor hopping Hamiltonian is
\begin{align}\label{eq:bandham1}
    H_{\bf i}^{\rm hop}&=-J_{\perp}\sum_{\alpha\in 1,2}\left(a_{{\bf i},\alpha}^{\dagger}a_{{\bf i}+1_y,\alpha}+a_{{\bf i},\alpha}^{\dagger}a_{{\bf i}+1_z,\alpha}+h.c.\right)\nonumber\\
     &\quad\quad\quad-\sum_{\alpha\in 1,2}J_{\alpha}\left(a_{{\bf i},\alpha}^{\dagger}a_{{\bf i}+1_x,\alpha}+h.c.\right).
  \end{align}
   The operators $a_{{\bf i},\alpha}^{\dagger}$ and $a_{{\bf i},\alpha}$ create and annihilate a particle in the Wannier function of unit cell ${\bf i}$ and band $\alpha=1,2$, respectively. The abbreviation $h.c.$ is the hermitian conjugate. The location ${\bf i}+1_x$ denotes the unit cell $(i_x+1,i_y,i_z)$. A similar notation is used for the other directions. The nearest-neighbor (NN) tunneling energy in band $\alpha$ along the $x$ axis is $J_{\alpha}=(1/M_x)\sum_{k_x}\cos(ka)\epsilon_{x,\alpha}(k_x)$, where the sum is over the $M_x$ allowed quasi-momenta $k_x$. Along the perpendicular directions the NN tunneling energy in the ground band is $J_{\perp}$. There is no cross-tunneling between separate bands as the Wannier functions are orthogonal. We have not included next-nearest neighbor tunneling since our numerical computation show that they are nearly two orders of magnitude smaller than the NN tunneling for the lattice depths on which we focus in this paper.

  The Hamiltonian $H_{\bf i}^{\rm cell}$ for unit cell ${\bf i}$ is 
  \begin{align}\label{eq:onsitebandham1}
    H_{\bf i}^{\rm cell}&= \frac{\delta}{2}\left(\hat n_{{\bf i},2}-\hat n_{{\bf i},1}\right)+\frac{1}{2}\sum_{\alpha\in 1,2}U_{\alpha}\hat n_{{\bf i},\alpha}(\hat n_{{\bf i},\alpha}-1)\nonumber\\
           &\quad\,+2U_{12}\hat n_{{\bf i},1}\hat n_{{\bf i},2}\nonumber\\
           &\quad\,      +\frac{1}{2}U_{12}\left(a_{{\bf i},1}^{\dagger}a_{{\bf i},1}^{\dagger}a_{{\bf i},2}a_{{\bf i},2}+a_{{\bf i},2}^{\dagger}a_{{\bf i},2}^{\dagger}a_{{\bf i},1}a_{{\bf i},1}\right),
  \end{align}
    where $\hat n_{{\bf i},\alpha}=a_{{\bf i},\alpha}^{\dagger}a_{{\bf i},\alpha}$ and the band-gap between bands $\alpha=1$ and $2$ is $\delta=(1/M_x)\sum_{k_x}\{\epsilon_{x,2}(k_x)-\epsilon_{x,1}(k_x)\}$. The on-site interaction terms $U_1$, $U_2$ and $U_{12}$ are obtained numerically from the local Wannier functions as $U_{\alpha}=g\int w_{{\bf i},\alpha}(\vec x)w_{{\bf i},\alpha}(\vec x)w_{{\bf i},\alpha}(\vec x)w_{{\bf i},\alpha}(\vec x)d\vec x$ and $U_{\alpha\beta}=g\int w_{\alpha}(\vec x)w_{\alpha}(\vec x)w_{\beta}(\vec x)w_{\beta}(\vec x)d\vec x$, where $g=4\pi\hbar^2a_s/m_a$ and $a_s$ is the $s$-wave scattering length. As a consequence of the nature of the Wannier functions, the term $U_{12}$ is comparable to $U_1$ and $U_2$ giving rise to strong density-density and pair-tunneling terms. It is the interplay between this pair-tunneling term and the band gap contribution that will enable us to achive large effective three-body interaction. 

For a symmetric double-well lattice, other atom-atom interaction terms between Wannier functions in bands $1$ and $2$ within a unit cell are strictly zero due to the parity of the Wannier functions. For small asymmetries, these other terms are small and can be neglected. Finally, we have not included terms due to atom-atom interactions between Wannier functions in neighboring unit cells, because they are an order of magnitude or more smaller than those in Eq.~\eqref{eq:onsitebandham1}.

Figures~\ref{fig:BHparameters} and \ref{fig:bandwannier} show parameters of the BH Hamiltonian as a function of lattice depth $V_0$ for ${}^{87}{\rm Rb}$ atoms at a fixed ratio of $V_1/V_0$ and fixed $V_2$. The plot shows that the interaction energies $U_1$, $U_2$ and $U_{12}$ are of equal importance, satisfy $U_2>U_{12}>U_1$, and are of the order of the band gap $\delta$. The tunneling energies are much smaller than the interaction energies. The tunneling energy $J_2$ for band $2$, as expected, is negative and $|J_2|>|J_1|$. Our choice of $V_2/E_R=40.0$ is such that, as shown in the next section, our Hubbard Hamiltonian has a superfluid region. 
\end{section}

\begin{section}{phase diagram using decoupling approximation}\label{sec:numericalMF}
  
  \begin{figure}
    \centering
    \includegraphics[width=3.45in]{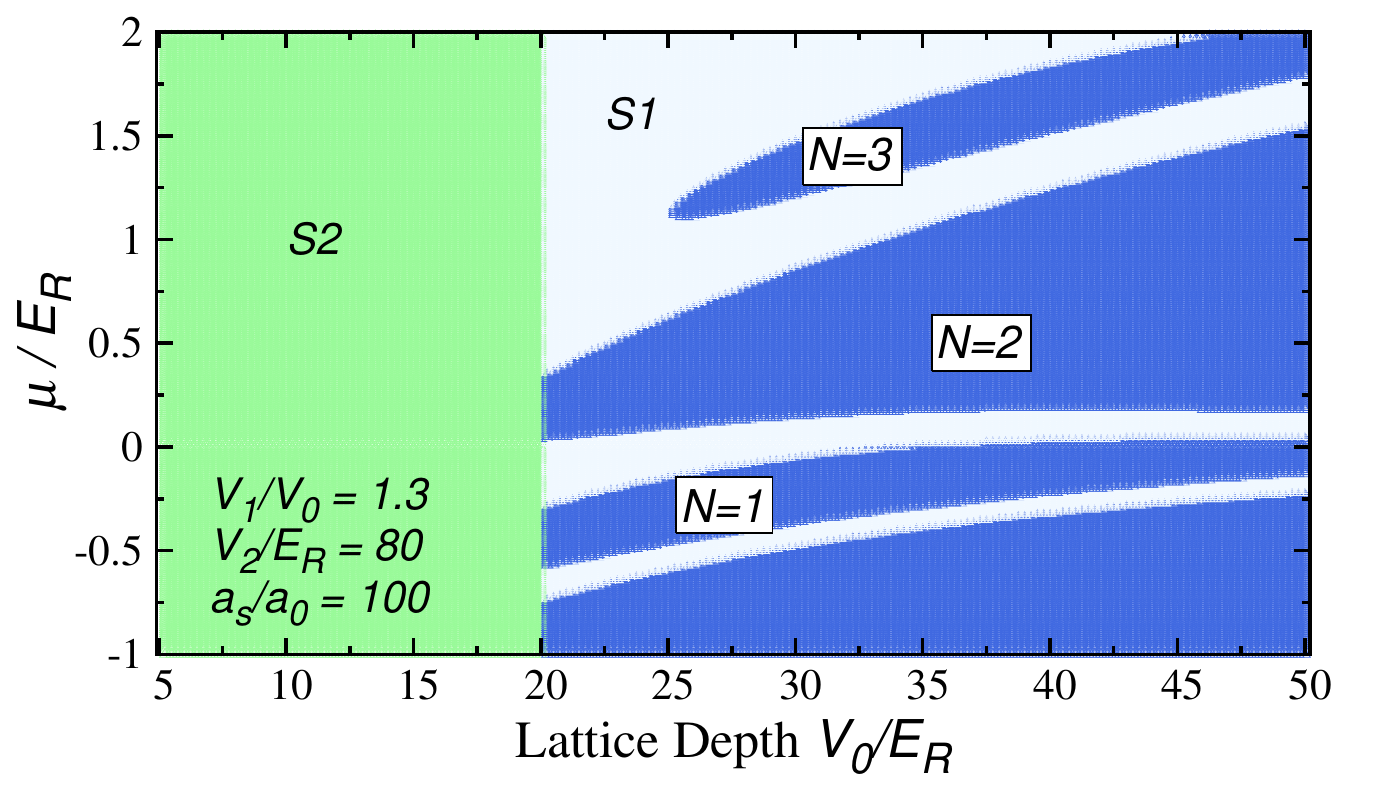}
    \caption{(color online) Phase diagram for ${}^{87}{\rm Rb}$ in a symmetric double-well lattice as a function of lattice depth $V_0$ in units of $E_R$ at $V_2/E_R=80.0$, $V_1/V_0=1.3$ and scattering length $a_s=100a_0$. The dark blue lobes are the Mott lobes for $N=1,2$ and $3$ atoms per unit cell. All other regions represent SF phases, with $S1$ denoting the case $\psi_1\ne 0,\psi_2=0$ and $S2$ denoting $\psi_2\ne 0$, $\psi_1=0$.} 
    \label{fig:numericalMF}
  \end{figure}
  
  In this section, we obtain the mean-field phase diagram for the BH Hamiltonian. A knowledge of the quantum phases as a function of lattice parameters is essential to prepare the system in the required many-body ground state. In addition, the results will be used in Sec.~\ref{sec:compareH} to discuss the validity of the effective Hamiltonian model. The phase diagram is obtained in the mean-field decoupling approximation \cite{van_oosten_quantum_2001}  by introducing a real-valued homogeneous superfluid order parameter for each band $\alpha=1,2$, i.e., $\psi_{\alpha}=\langle a_{{\bf i},\alpha}^{\dagger}\rangle=\langle a_{{\bf j},\alpha}\rangle$ and approximate $a_{{\bf i},\alpha}^{\dagger}a_{{\bf j},\alpha}$ by $\psi_{\alpha}\left(a_{{\bf i},\alpha}^{\dagger}+a_{{\bf j},\alpha}\right)-\psi_{\alpha}^2$ for ${\bf i}\ne{\bf j}$. In the grand-canonical ensemble, this leads to the grand-canonical potential or Hamiltonian $H^{\rm MF}=\sum_{\bf i}(H_{{\bf i}}^{\rm MF}-\mu a_{{\bf i},\alpha}^{\dagger}a_{{\bf i},\alpha})$, where
  \begin{align}\label{eq:MFHamiltonian}
    H_{\bf i}^{\rm MF}=&H_{\bf i}^{\rm cell}-\sum_{\alpha\in 1,2}\left(a_{{\bf i},\alpha}^{\dagger}+a_{{\bf i},\alpha}\right)\left(2J_{\alpha}+z_{\perp}J_{\perp}\right)\psi_{\alpha}\nonumber\\
      \quad&+\sum_{\alpha\in 1,2}\left(2J_{\alpha}+z_{\perp}J_{\perp}\right)\psi_{\alpha}^2
  \end{align}
  and $z_{\perp}$ is the number of neighboring unit cells along the perpendicular directions. In our case, $z_{\perp}=4$. The ground state of the grand-canonical potential $H^{\rm MF}$ at a given chemical potential $\mu$ corresponds to order parameters $(\psi_1,\psi_2)$ that minimize its ground state energy. We have performed these calculations in a basis of Fock states $|n_1,n_2\rangle$, where $n_{\alpha}$ is the number of atoms in band $\alpha$ in an unit cell. We reach numerical convergence for $n_1,n_2\le N_{\rm max}=10$. A Mott insulator (MI) state corresponds to $\psi_1=\psi_2=0$, while a superfluid (SF) state is obtained when either $\psi_1,\psi_2\ne 0$.

  Figure ~\ref{fig:numericalMF} shows the mean-field phase diagram for our symmetric double-well lattice. The figure shows Mott lobes up to $N=3$ atoms per unit cell. We find that the Mott lobes with odd $N$ are reduced in size compared to adjacent lobes with even $N$, a feature also noted and discussed in \cite{zhou_condensates_2011}. For the SF phase $S1$, the order parameter $\psi_1\ne 0$, but $\psi_2=0$, while for the SF phase $S2$ $\psi_2\ne 0$ but $\psi_1=0$. Both the order parameters have a jump at the $S1$-$S2$ phase-boundary corresponding to a first-order phase transition. For the relatively shallow depths used in all directions in Fig.~\ref{fig:numericalMF}, superfluidity extends over all dimensions. For larger depths along either $x$ or $y$ and $z$, the superfluid can be localized to fewer dimensions. A further stability analysis around the mean-field solution is then required to identify this distinction \cite{fisher_1973,smerzi_2002,altman_2005}. Such knowledge, however, is not crucial for deriving the effective Hamiltonian model.

  A cutting off of the Mott lobes for $N=1$ and $2$ at the phase-boundary to phase $S2$ is apparent in Fig.~\ref{fig:numericalMF}. From a separate analytical calculation of the phase-boundary of $H^{\rm MF}$, excluding the interaction induced pairing terms, we find that this cut-off occurs when $2J_{2}+z_{\perp}J_{\perp}=0$. For the lattice parameters used in Fig.~\ref{fig:numericalMF} this occurs when $V_0/E_R\approx 20$, consistent with our numerical simulations including the pairing term. 

The lattice parameters in Fig.~\ref{fig:numericalMF} were particularly chosen so as to have a good representation of all the three quantum phases. Since our primary aim in this paper is to show the presence of large effective three-body interactions in double-well optical lattices, we do not provide phase diagrams for a wider set of lattice parameters. Instead, we note that in our case, we are mainly interested in the phase $S1$ and the corresponding SF-Mott phase-boundary. Also, the region $S1$ grows relative to those of other phases as we decrease the lattice depth along the perpendicular directions. 
\end{section}

\begin{section}{effective hamiltonian}
  \label{sec:Heff}
  
  We now show that the low energy states of the multi-band BH Hamiltonian defined in Eqs.~\eqref{eq:bandham1} and \eqref{eq:onsitebandham1} can be obtained from an effective Hamiltonian $H_{\rm eff}$ that has strong three-body interactions. Constructing $H_{\rm eff}$ is a three-step process. The first step is discussed in Sec.~\ref{sec:MPE} and involves diagonalizing the on-site Hamiltonian $H_{\bf i}^{\rm cell}$ in unit cell $\bf i$  to obtain many-particle (MP) energy levels. Using these MP levels an effective on-site interaction Hamiltonian is constructed in Sec.~\ref{sec:effHint}. Finally, in Sec.~\ref{sec:effHhop} we calculate the effective tunneling Hamiltonian that couples the MP states of adjacent unit cells.

\begin{subsection}{Many-particle energy levels}
  \label{sec:MPE}
  
  \begin{figure}
  \subfloat{\includegraphics[width=3.2in]{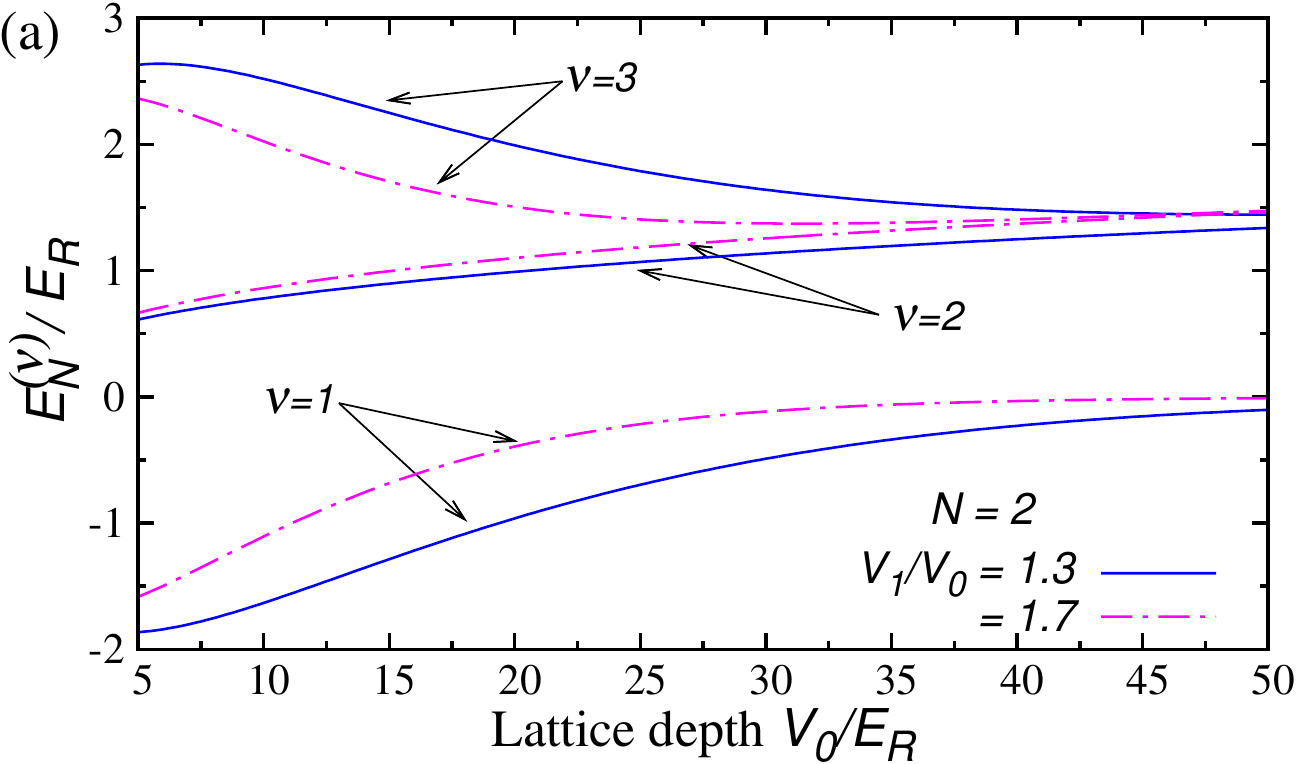}\label{fig:MPE2}}
  \hspace{9mm}
  \subfloat{\includegraphics[width=3.2in]{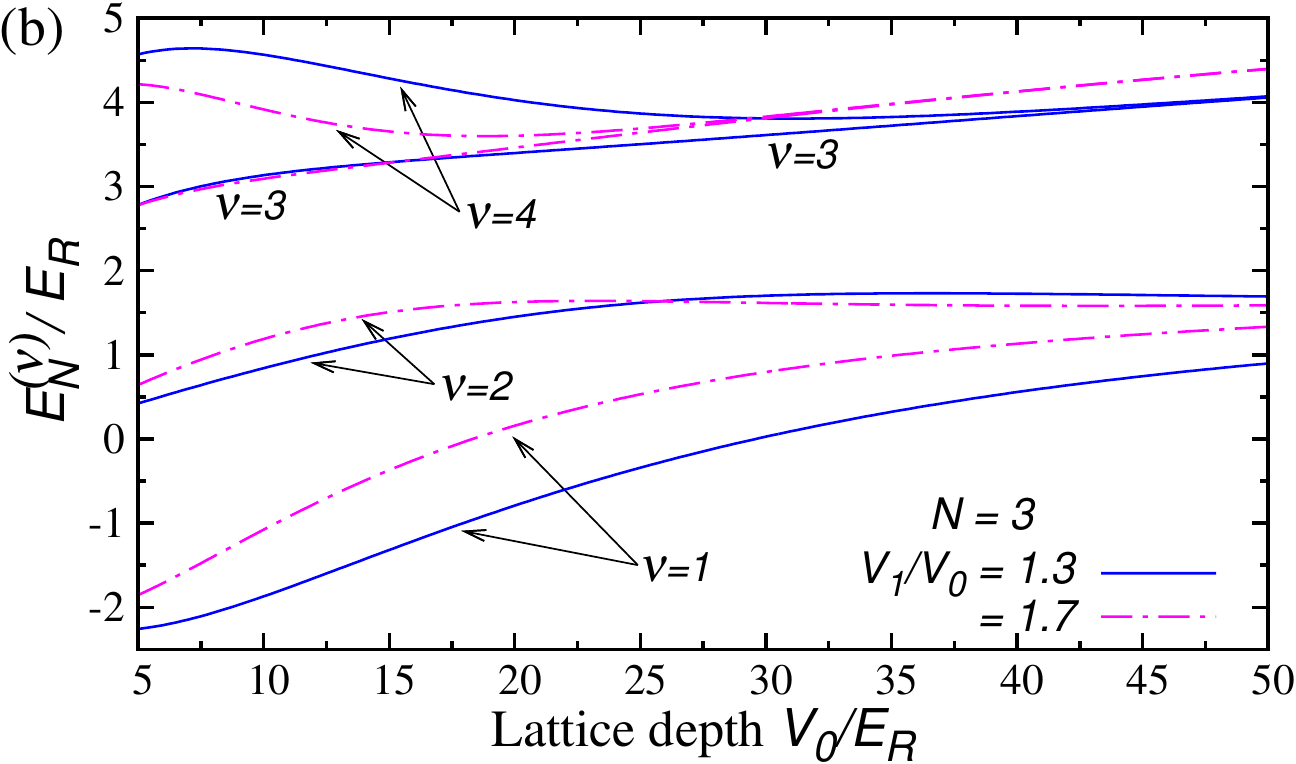}\label{fig:MPE3}}
  \caption{(color online)  (a) Many-particle energy levels ${\cal E}_N^{(\nu)}$ for two atoms per unit cell, as functions of lattice depth $V_0$ in units of $E_R$. The solid blue lines are for $V_1/V_0=1.3$, while the dot-dashed  magenta curves are for $V_1/V_0=1.7$.  (b) Similar plot for three atoms per unit cell. Plots are for ${}^{87}$Rb in a symmetric lattice with $V_2/E_R=40.0$ and scattering length $a_s=100a_0$.}
\end{figure}

 We are interested in lattice parameters for which the average band gap $\delta$ and interaction energies $U_1,U_2$ and $U_{12}$ are much larger than the tunneling energies $J_1,J_2$ and $J_{\perp}$.  We thus diagonalize the on-site Hamiltonian $H_{\bf i}^{\rm cell}$ in the Fock state basis $|n_1,N-n_1\rangle$ for $N$ atoms in unit cell ${\bf i}$ with $n_1$ atoms in band $\alpha=1$. This gives the many-particle (MP) eigen energies ${\cal E}_N^{(\nu)}$ and eigen states $|\nu,N\rangle$ with $\nu=\{1,\ldots,N+1\}$. We refer to these $\nu$ states as vibrational states. The MP eigen-states in terms of the Fock states are
  \begin{align}\label{eq:MPstates}
    |\nu,N\rangle = \sum_{n_1=0}^{N}{\cal C}_{n_1}^{(\nu)}(N)|n_1,N-n_1\rangle,
  \end{align}
  with real coefficients ${\cal C}_n^{(\nu)}(N)$. It is good to note that the many-body ground state of the Hamiltonian in Eq.~\eqref{eq:bandham1} in a Mott lobe is a product of $|\nu,N\rangle$ states with $\nu=1$, one for each site.

For $N=1$ the eigen states are simply the Fock states $|1,0\rangle$ and $|0,1\rangle$ with energy ${\cal E}_1^{(1)}=-\delta/2$ and ${\cal E}_1^{(2)}=\delta/2$. A plot of the band gap $\delta$ as a function of lattice depth $V_0$ is shown in Fig.~\ref{fig:BHparameters}. Figures~\ref{fig:MPE2} and \ref{fig:MPE3} show the MP energy levels ${\cal E}_N^{(\nu)}$ as a function of lattice depth $V_0$ and at fixed $V_1/V_0$ and $V_2$ for two and three atoms, respectively. Their behavior can be understood by noting that the band gap $\delta$ decreases with increasing lattice depth $V_0$, and for larger depths $\delta/U\ll 1$ with $U\approx U_{1}\approx U_{2}\approx U_{12}$. Consequently, we can show using perturbation theory that for $N=2$, ${\cal E}_2^{(1)}=-\delta^2/(2U)$, ${\cal E}_2^{(2)}=2U$ and ${\cal E}_2^{(3)}=2U+\delta^2/(2U)$. For $N=3$ atoms, ${\cal E}_3^{(1)}=2U-\delta-3\delta^2/6U$, ${\cal E}_3^{(2)}=2U+\delta-3\delta^2/6U$, and ${\cal E}_3^{(3)}={\cal E}_3^{(4)}=6U+3\delta^2/6U$. In conjunction with Fig.~\ref{fig:bandwannier} these expressions give a fair estimate of ${\cal E}_N^{(\nu)}$ for large $V_0$. We also note that the energy of the ground-vibrational $\nu=1$ state is well separated from that of the state $\nu=2$ with a spacing much larger than $J_1$, $J_2$ or $J_{\perp}$. For the effective Hamiltonian picture, we assume that the atoms only populate the $\nu=1$ states.

We should further note that interactions in Eq.~\eqref{eq:onsitebandham1} mix the Fock states $|n_1,N-n_1\rangle$ and the vibrational ground states $|\nu=1,N\rangle$ at large lattice depths have considerable contribution from Fock states with atom population in the first excited band.
\end{subsection}

\begin{subsection}{Effective interaction Hamiltonian}
  \label{sec:effHint}
  
  \begin{figure}
    \centering
    \includegraphics[width=3.2in]{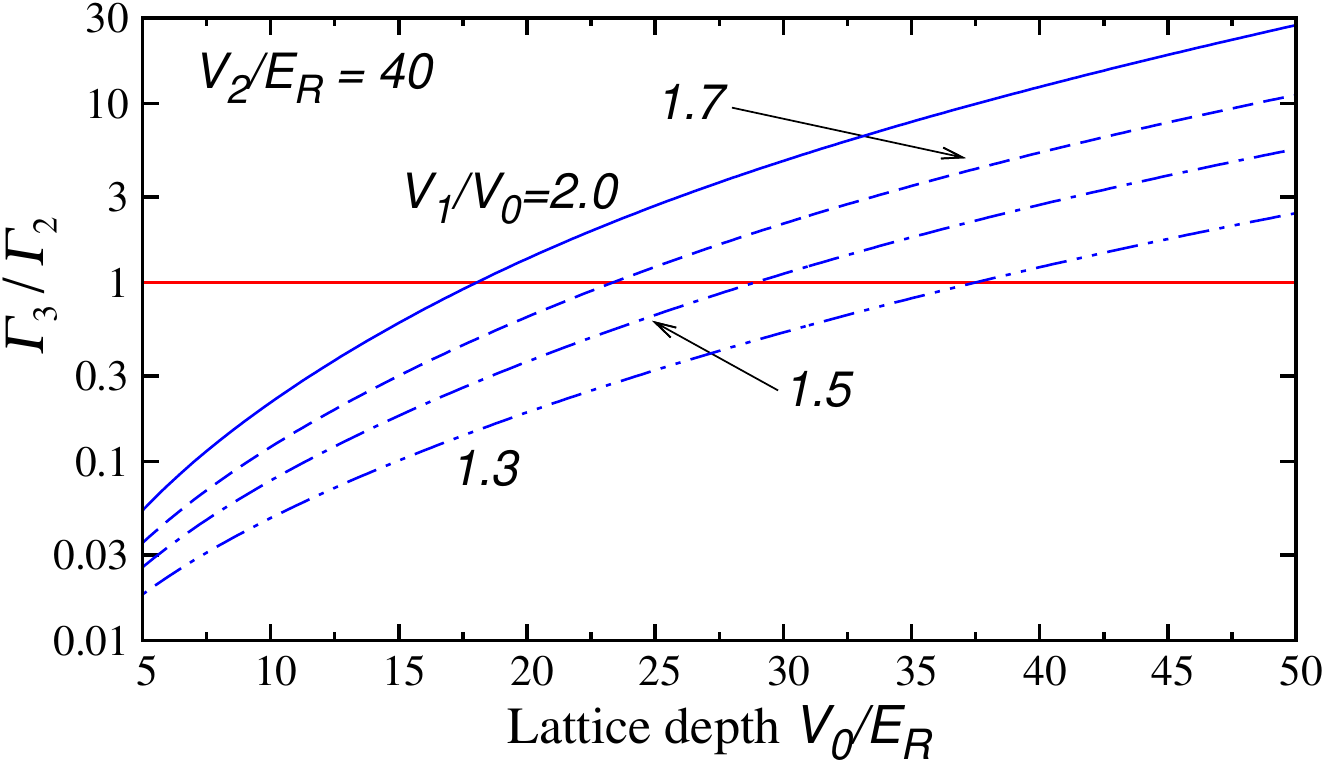}
    \caption{(color online) Log-linear plot of the ratio of three- to two-body interaction strength $\Gamma_3/\Gamma_2$ as a function of lattice depth $V_0$ in units of $E_R$ for various ratios of $V_1/V_0$. The horizontal red line denotes $\Gamma_3/\Gamma_2=1$. The plot is for ${}^{87}$Rb in a symmetric lattice at $V_2/E_R=40.0$ and scattering length $a_s=100a_0$.}
    \label{fig:effHint}
  \end{figure}

  We can now construct the effective on-site interaction Hamiltonian similar to Eq.~\eqref{eq:modelH} in unit cell ${\bf i}$ based on the large energy separation between the $\nu=1$ and $2$ vibrational states for each $N$. In other words we assume that the atoms only populate the ground vibrational states with energies ${\cal E}_N^{(1)}$ that are reproduced by the effective on-site interaction Hamiltonian
  \begin{align}
    \label{eq:effHint}
    H_{\rm eff}^{\rm int}&=\sum_{\bf i}\sum_{m=1}^{3}\frac{1}{m!}\Gamma_m b_{\bf i}^{\dagger m} b_{\bf i}^m,
  \end{align}
 where $b_{\bf i}$ and $b_{\bf i}^\dagger$ are effective bosonic annihilation and creation operators for unit cell $\bf i$ and state $\nu=1$. The coefficients $\Gamma_m$ are the $m-$body interaction strengths, which we restrict to $m\le 3$, and are found by mapping ${\cal E}_N^{(1)}$ to the eigen-energies of $H_{\rm eff}^{\rm int}$. Some algebra shows $\Gamma_1={\cal E}_1^{(1)}$, $\Gamma_2={\cal E}_2^{(1)}-2{\cal E}_1^{(1)}$ and $\Gamma_3={\cal E}_3^{(1)}-3({\cal E}_2^{(1)}-{\cal E}_1^{(1)})$.

 Figure~\ref{fig:effHint} shows $\Gamma_3/\Gamma_2$ as a function of lattice depth $V_0$ for various $V_1/V_0$. We observe that $\Gamma_3/\Gamma_2$ increases with $V_0$ and $V_1/V_0$, i.e. with increasing lattice depth. In fact, choosing appropriate lattice parameters can produce a three-body strength that is larger than the two-body one. Using the same approximations as used in Sec.~\ref{sec:MPE} for large lattice depths, we find $\Gamma_2\approx \delta-\delta^2/(2U)+\mathcal{O}(\delta^3)$ and $\Gamma_3\approx 2U-2.5\delta+1.3\delta^2+\mathcal{O}(\delta^3)$. Consequently, $\delta$ and $\Gamma_2\to 0$ while $\Gamma_3$ remains finite for larger lattice depths and $\Gamma_3\gg\Gamma_2$. 

For an asymmetric double-well lattice with $k_Lb\ne\pi/4$, $\delta$ and equivalently $\Gamma_2$ depend on the asymmetry and remain finite for a large lattice depth. A symmetric lattice is therefore the most favorable case with which to reach sizeable $\Gamma_3.$ We should also note that it is not possible to obtain $\Gamma_2\ll\Gamma_3$ for a single-well lattice since the band gap increases with lattice depth.
\end{subsection}

\begin{subsection}{Effective tunneling Hamiltonian}
  \label{sec:effHhop}

  In this section we study the coupling between the MP states $|\nu,N\rangle$ of neighboring unit cells. Since tunneling involves two adjacent unit cells, we label states as $|\nu,N\rangle_{\bf i}|\nu',M\rangle_{{\bf i}+1}$, where $N$ and $M$ are the atom numbers in unit cell $\bf i$ and ${\bf i}+1_x$, ${\bf i}+1_y$ or ${\bf i}+1_z$, respectively.  As in the previous subsection, we restrict our discussion to the $\nu=1$ ground state and simply write the initial and final states as $|N,M;{\bf i},{\bf i}+1\rangle\rangle \equiv|\nu=1,N\rangle_{\bf i}|\nu'=1,M\rangle_{{\bf i}+1}$.

\begin{figure}
  \centering
  \subfloat{\includegraphics[width=3.2in]{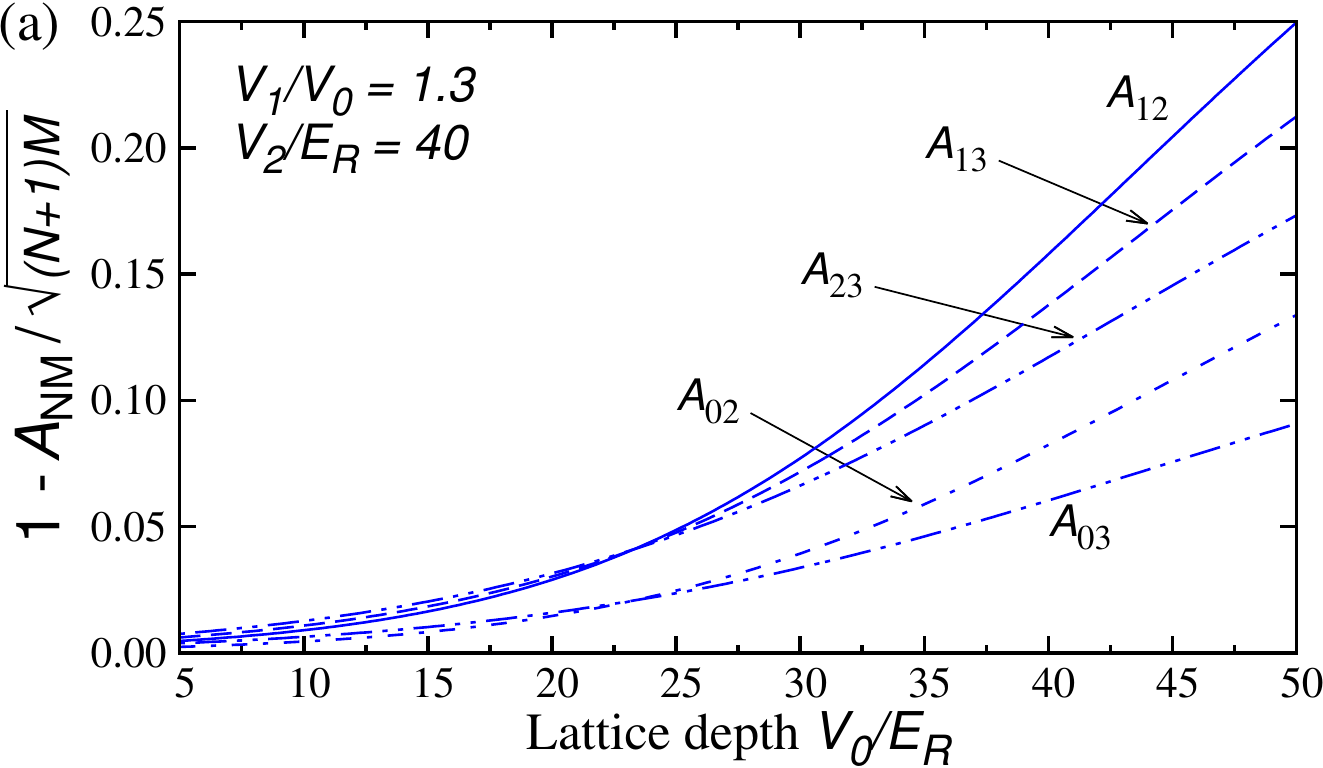}\label{fig:ANM}}\\
  \subfloat{\includegraphics[width=3.2in]{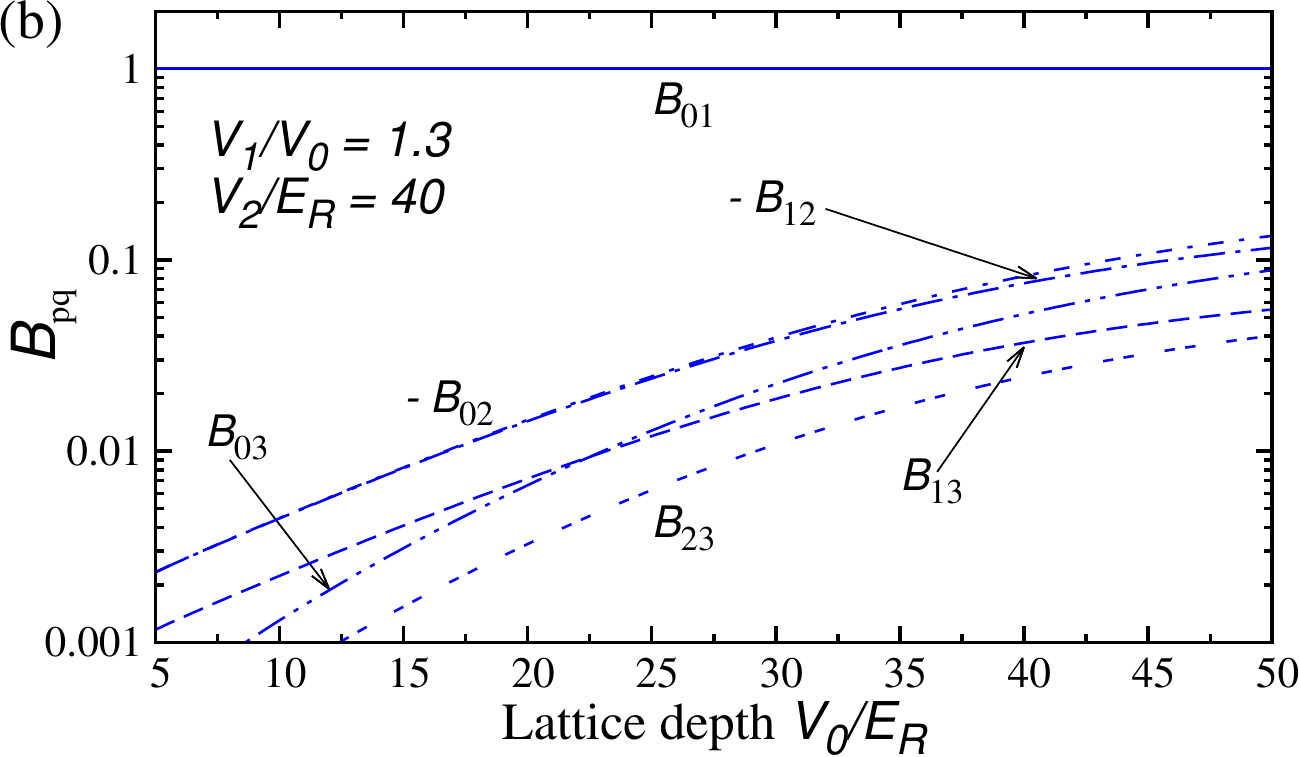}\label{fig:BNM}}
  \caption{  (a) Plot of ratio $1-{\cal A}_{NM}/\sqrt{(N+1)M}$ as a function of lattice depth $V_0$. The coefficient ${\cal A}_{01}$ is not shown because the corresponding ratio is zero. (b) Plot of ${\cal B}_{pq}$ as a function of lattice depth $V_0$. Both the panels are for ${}^{87}$Rb with $k_Lb=\pi/4$, $V_1/V_0=1.3$, $V_2/E_R=40.0$ and $a_s=100a_0$.}
\end{figure}

  With these definitions we can express tunneling between neighboring unit cells along the $x$ axis as
  \begin{eqnarray}\label{eq:mpehop1}
   \lefteqn{ H_{x,\rm eff}^{\rm hop}=-J_1\sum_{\bf i}{\sum_{N,M}}^{'}}\\
    &&\Big\{{\cal A}_{NM}|N\!+\!1,M\!-\!1;{\bf i},{\bf i}+1_x\rangle\rangle\langle\langle N,M;{\bf i},{\bf i}+1_x|+h.c.\Big\}\nonumber
  \end{eqnarray}
  where the sums over $N$ and $M$ are restricted to $0\le N<M$, and using Eqs.~\eqref{eq:bandham1} and \eqref{eq:MPstates}, 
   \begin{align}\label{eq:ANM}
     -J_1{\cal A}_{NM}&=\langle\langle N\!+\!1,M\!-\!1;{\bf i},{\bf i}+1_x|H_{\bf i}^{\rm hop}|N,M;{\bf i},{\bf i}+1_x\rangle\rangle\nonumber\\
     &\approx -J_1\sum_{n=0}^N\sum_{m=1}^M\sqrt{(n+1)m}\,{\cal C}_n^{(1)}(N)\,{\cal C}_{n+1}^{(1)}(N+1)\nonumber\\
     &\quad\quad \quad\times\,{\cal C}_m^{(1)}(M)\,{\cal C}_{m-1}^{(1)}(M+1),
   \end{align}
are unit-cell-independent dimensionless real coefficients. Here, we only include tunneling through the term proportional to $J_1$ in Eq.~\eqref{eq:bandham1}, since the $\nu=1$ vibrational level is predominantly determined by the $|N,0\rangle$ Fock state. Thus, the contribution from tunneling through the $J_2$ term is negligible. We limit ourselves to $N,M\le 3$, i.e. up to three atoms per site, and only require the six independent ${\cal A}_{01},\,{\cal A}_{02},\,{\cal A}_{12},\,{\cal A}_{03},\,{\cal A}_{13}$ and ${\cal A}_{23}$. In Fig.~\ref{fig:ANM} we plot the ratio $1-{\cal A}_{NM}/\sqrt{(N+1)M}$ for these six coefficients as a function of lattice depth $V_0$. The expression $\sqrt{(N+1)M}$ corresponds to the coefficient expected for a simple single-well BH Hamiltonian. The coefficients ${\cal A}_{NM}$ deviate only slightly from these values for $V_0/E_R<30$ in Fig.~\ref{fig:ANM}. More generally, we find that the correction becomes larger for larger lattice depth. 

The effective tunneling Hamiltonian $H_{x,\rm eff}^{\rm hop}$ has atom-number dependent parameters. We find it convenient to write $H_{x,\rm eff}^{\rm hop}$ in an alternative form as a sum of multi-body hopping operators 
  \begin{align}\label{eq:effHhopx1}
    H_{x,\rm eff}^{\rm hop} &= -J_1\sum_{\bf i}{\sum_{q,p}}^{'}\Big\{{\cal B}_{pq}\,\left(b_{\bf i}^{\dagger}\right)^{p+1}\,\left(b_{{\bf i}+1_x}^{\dagger}\right)^{q-1}\Big.\nonumber\\
      &\quad\quad\quad\quad\quad\quad\quad\quad \,\Big.\times\left(b_{\bf i}\right)^p\,\left(b_{{\bf i}+1_x}\right)^{q} + h.c.\Big\},
  \end{align}
  where the sums over $p$ and $q$ are restricted to $0\le p<q$. An $m$-body hopping operator corresponds to terms where $p+q=m$. The six coefficients ${\cal B}_{pq}$ are found from solving the linear system
  \begin{align}\label{eq:BNM}
    & {\cal A}_{NM}  = \sum_{q=1}^3\sum_{p=0}^{q-1}{\cal B}_{pq}{\cal F}(N,M,p,q)\\
    \text{with}\quad &{\cal F}(N,M,p,q) = \sqrt{(N+1)M}\frac{(M-1)!N!}{(M-q)!(N-p)!}.\nonumber
  \end{align}
 
  Figure ~\ref{fig:BNM} shows a plot of ${\cal B}_{pq}$ as a function of lattice depth $V_0$. The most important term is ${\cal B}_{01}=1$, the usual single-particle tunneling term between adjacent unit cells. The next two leading terms are the two- and three-body operators with strength ${\cal B}_{02}$ and ${\cal B}_{12}$, respectively. Their values are negative. Other terms are a factor of two or more smaller and positive. 

 Figure~\ref{fig:MPE3} shows that the energy separation between the $\nu=1$ and $2$ vibrational states for $N=3$ atoms decreases with increasing ratio $V_1/V_0$. This might seem to invalidate the assumption that the atoms only populate the ground vibrational state for larger lattice depths. For such a scenario with large $V_1/V_0$, however, we can choose a smaller lattice depth $V_0$ to ensure that the ground vibrational state is well separated, and still have a large $\Gamma_3>\Gamma_2$. Moreover, the coupling between the ground and excited vibrational states in neighboring unit cells are of the order of the tunneling energies with strength proportional to $J_1^2/\sqrt{U^2+\delta^2}$, which decrease with increasing $V_1/V_0$, and stay much smaller than the  energy separation between the ground and excited vibrational states.

The effective tunneling Hamiltonians along perpendicular directions are obtained by replacing $J_1$ by $J_{\perp}$. The numerical values for the coefficients ${\cal B}_{pq}$ are the same as those along the $x$ axis. This finally leads to the effective tunneling Hamiltonian $H_{\rm eff}^{\rm hop}=H_{x,\rm eff}^{\rm hop}+H_{y,\rm eff}^{\rm hop}+H_{z,\rm eff}^{\rm hop}$ and the total effective Hamiltonian $H_{\rm eff}=H_{\rm eff}^{\rm int}+H_{\rm eff}^{\rm hop}$.
\end{subsection}
\end{section}

\begin{section}{validity of the effective Hamiltonian}
  \label{sec:compareH}

  \begin{figure}
    \centering
    \subfloat{\includegraphics[width=3.2in]{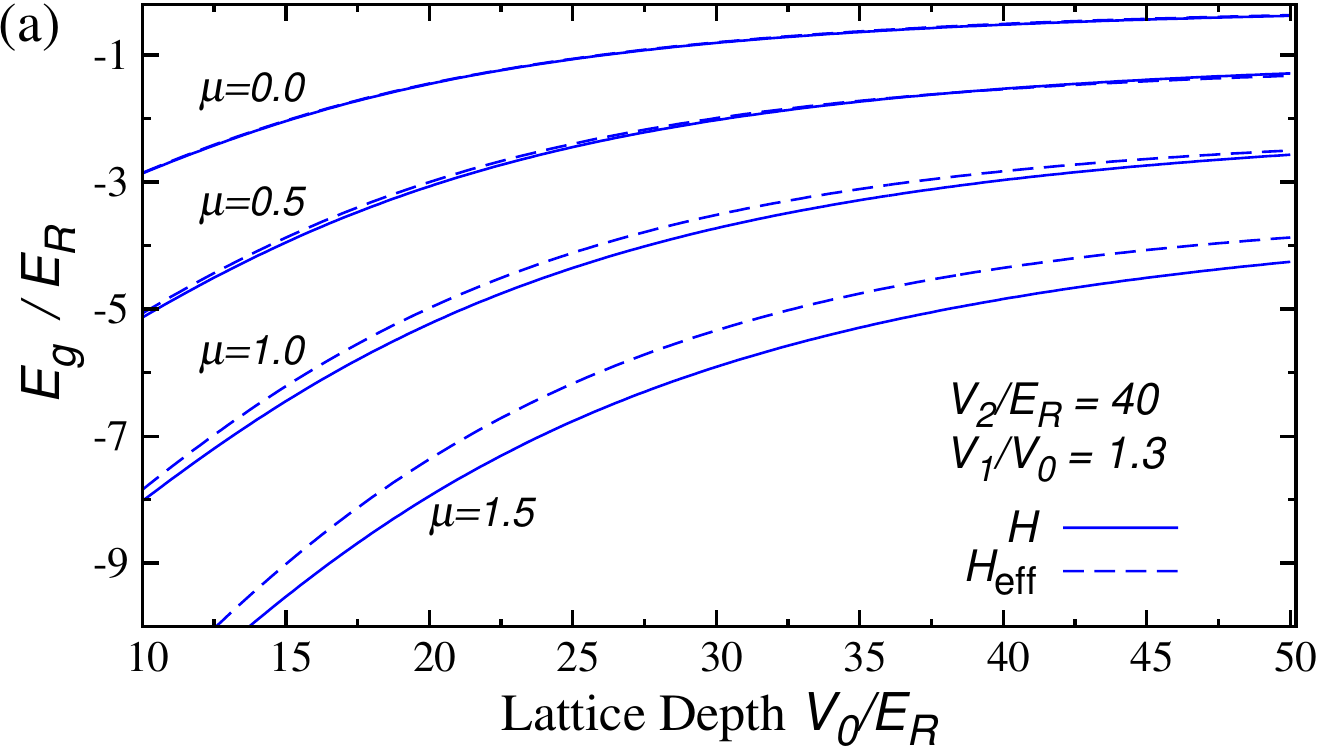}\label{fig:compareEg}}\\
    \subfloat{\includegraphics[width=3.45in]{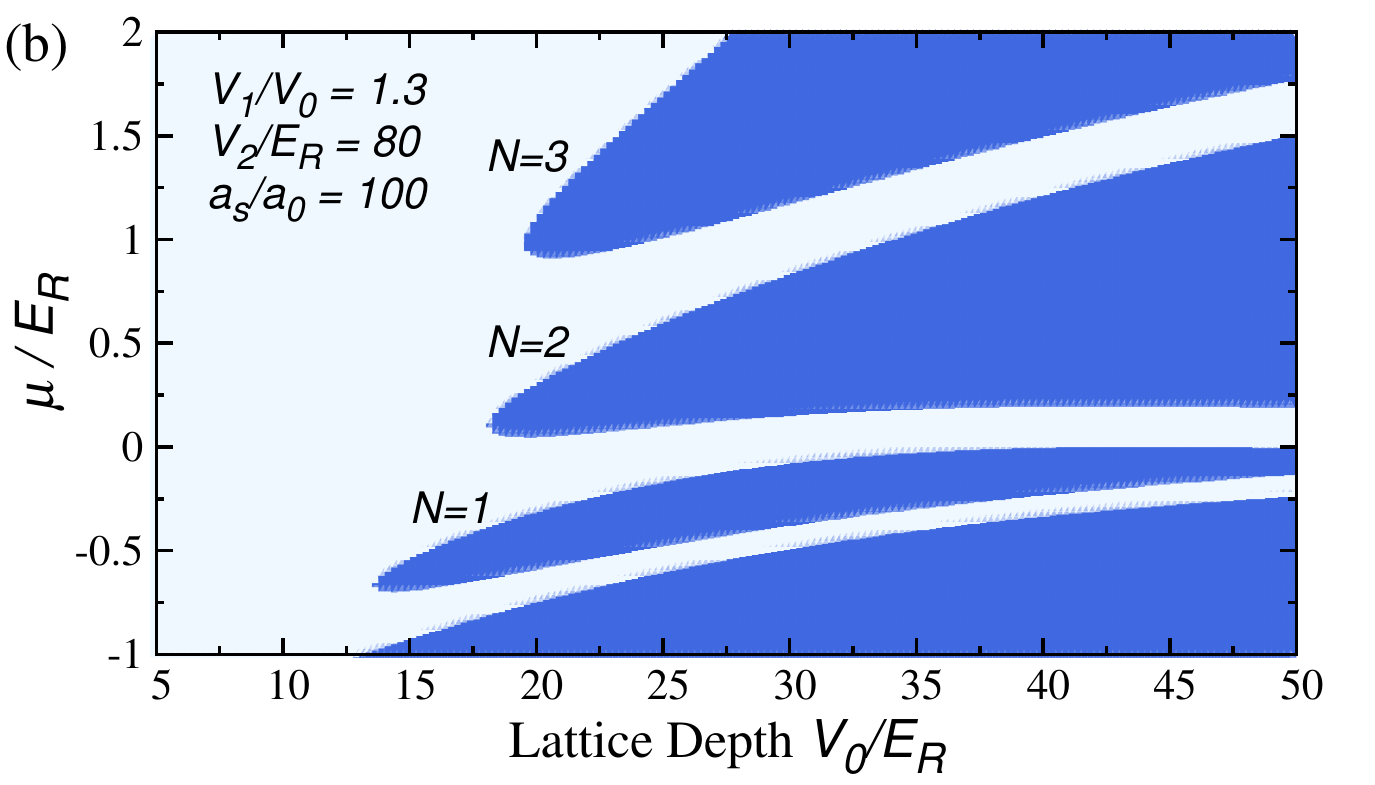}\label{fig:HeffMF}}
    \caption{(color online)  (a) Plot of the mean-field ground-state energy $E_g$ as a function of lattice depth $V_0$ in units of $E_R$ and $V_2/E_R=40$ for different chemical potentials $\mu$. The solid lines represent the results for the full Hamiltonian, while the dashed lines show those for the effective Hamiltonian. Lattice parameters are chosen such that the system is in the SF region $S1$. (b) Mean-field SF-Mott phase diagram for $H_{\rm eff}$ as a function of lattice depth $V_0/E_R$ at $V_2/E_R=80$. The deep blue areas are Mott lobes surrounded by a SF region. Other parameters are the same as those used in Fig.~\ref{fig:numericalMF}}.
  \end{figure}

  In this section, we discuss the validity of the effective Hamiltonian developed in Sec.~\ref{sec:Heff}. It was obtained by assuming that the atoms only occupy the ground vibrational ($\nu=1$) states and coupling to excited $\nu$ states is negligible. This assumption is justified as the energy gap between the ground and excited vibrational states (see Figs.~\ref{fig:MPE2} and \ref{fig:MPE3}) is much larger than the coupling strengths, which are of the order of the tunneling energies. We now verify this assumption further by comparing mean-field phase diagrams.

   For our purposes it is sufficient to further simplify $H_{\rm eff}$ to
  \begin{align}\label{eq:effH1}
    H_{\rm eff}&=\sum_{\bf i}\left\{\sum_{m=1}^{3}\frac{1}{m!}\Gamma_m b_{\bf i}^{\dagger m} b_{\bf i}^m-J_1b^{\dagger}_{\bf i}b_{{\bf i}+1_x}\right.\nonumber\\
    &\quad\,\left.-J_{\perp}\left(b^{\dagger}_{\bf i}b_{{\bf i}+1_y}+b^{\dagger}_{\bf i}b_{{\bf i}+1_z}\right)+h.c.\right\},
  \end{align}
  where we have only kept the leading-order single-body tunneling terms in Eq.~\eqref{eq:effHhopx1}. Thus, $H_{\rm eff}$ is a single-band BH Hamiltonian with an added effective three-body term proportional to $\Gamma_3$. Similar to Sec.~\ref{sec:numericalMF}, we introduce a real order parameter $\psi=\langle b_{\bf i}\rangle=\langle b_{\bf i}^{\dagger}\rangle$ and determine the ground-state energy, $E_g$, for the corresponding mean-field Hamiltonian in the grand-canonical ensemble by self-consistently minimizing the energy with respect to the order parameter $\psi$. 

We first confirm that $H_{\rm eff}$ is a good model for lattice parameters where the full Hamiltonian in Eqs.~\eqref{eq:bandham1} and \eqref{eq:onsitebandham1} has a superfluid ground state in the $S1$ region. Figure~\ref{fig:compareEg} compares the ground-state energy $E_g$ of the full mean-field and the effective Hamiltonian as a function of lattice depth $V_0$ for various values of $\mu$. This graph is computed for $V_2/E_R=40$, half the size used in Fig.~\ref{fig:numericalMF}. The mean-field ground state of the full Hamiltonian is then in the $S1$ region, with order parameter $\psi_2=0$, for the entire parameter space $(\mu,V_0)$ shown in Fig.~\ref{fig:compareEg}. Similarly, the ground state of the effective Hamiltonian is superfluid. The mean atom number per unit cell for the four $\mu$s shown in Fig.~\ref{fig:compareEg} are $\bar n=1.56$, $2.37$, $3.19$ and $4.01$  at a lattice depth of $V_0/E_R=30$, respectively.

 The ground-state energies for the two models show similar trends, have differences that are nearly independent of $V_0$ and agree better for smaller chemical potentials. The larger discrepancy for larger $\bar n$ is not due to virtual excitations to higher $\nu$ vibrational states or $m>1$ hopping terms since $J_1{\cal B}_{pq}$ or $J_{\perp}{\cal B}_{pq}$ with $p,q\ne 0,1$ are much smaller than the recoil energy. For $\bar n>3$ we need to include four- or higher-body interaction terms to get better agreement.

  Next, we study whether $H_{\rm eff}$ is valid as we approach the SF-Mott phase-boundary. Unlike the ground state energy, the mean-field phase boundary can be analytically obtained within second-order perturbation theory \cite{van_oosten_quantum_2001}, treating $-(2J_1+z_{\perp}J_{\perp})\left(b_{\bf i}^{\dagger}+b_{\bf i}\right)\psi$ as a perturbation, and is given by 
  \begin{align}
    \label{eq:a2}
    0 &= J_{\rm eff}\left\{ 1 + \frac{(g+1)J_{\rm eff}}{-\Gamma_1+\frac{1}{2}g(\Gamma_3-2\Gamma_2)-\frac{1}{2}g^2\Gamma_3+\mu}+\right.\nonumber\\ 
    &\quad\left.\frac{gJ_{\rm eff}}{(\Gamma_3-\Gamma_2+\Gamma_1)+\frac{1}{2}g(2\Gamma_2-3\Gamma_3)+\frac{1}{2}g^2\Gamma_3-\mu}\right\},\nonumber
  \end{align}
  with $J_{\rm eff}=(2J_1+z_{\perp}J_{\perp})$ and the Fock state $|g\rangle$ is the zeroth-order ground state of the grand-canonical potential. Figure~\ref{fig:HeffMF} shows the corresponding SF-Mott phase diagram for $V_2/E_R=80$ and can thus be directly compared with the phase diagram shown in Fig.~\ref{fig:numericalMF} for the full Hamiltonian. For $V_0/E_R>20$ the Mott lobes for $N=0$, $1$ and $2$ are nearly identical. The $N=3$ lobe for the effective Hamiltonian, however, is significantly larger than that of the full Hamiltonian. In fact, the tip of the Mott lobe has shifted to smaller lattice depth.  We have to include $\Gamma_4$ to correctly model this lobe. Finally, our effective Hamiltonian does not describe $S2$ for $V_0/E_R<20$. In this phase, the full Hamiltonian has a non-zero order parameter in the second band. Our effective Hamiltonian assumes no population in the second band and can not represent these states.  
\end{section}

\begin{section}{conclusion}
  \label{sec:conclusion}
  We have shown that the low energy states of a system of trapped atoms in a double-well optical lattice can emulate a Hubbard model with strong three-body interactions. The optical lattice with a double-well potential along one direction and single period lattice along the remaining directions has two nearly degenerate bands. The Hamiltonian has a strong pair-tunneling contribution between the two Wannier functions within a unit cell. The interplay between this interaction and the band gap plays an important role in determining the behavior of the system. In particular, we find that the low energy states of such a system are quite accurately described by an effective single-band Hamiltonian with a strong three-body interaction, whose strength can be easily tuned by changing the lattice parameters. Surprisingly, tunneling between the neighboring unit cells in the effective Hamiltonian has, to good approximation, the same structure as that for a particle hopping in a single-band BH model. By comparing the numerically obtained ground state energy and phase diagram of the full and effective Hamiltonians, we verified that the effective Hamiltonian model is an excellent approximation over a wide range of lattice parameters, both in the superfluid and Mott insulator phases.
\end{section}

\bibliography{ref}
\end{document}